\documentclass[aps,prd,showkeys,superscriptaddress,twocolumn,nofootinbib,floatfix]{revtex4-1}

\usepackage{amsmath}
\usepackage{amssymb}
\usepackage{amsthm}
\usepackage{mathrsfs}
\usepackage{graphicx}
\usepackage{epstopdf}
\usepackage{fancyhdr}
\usepackage{array}
\usepackage[all]{xy}
\usepackage{eufrak}
\usepackage{euscript}
\usepackage{enumerate}
\usepackage{slashed}
\usepackage{hyperref}
\usepackage{subfigure} 
\usepackage{epstopdf} 

\hypersetup{pdftex,colorlinks=true,linkcolor=blue,citecolor=blue,menucolor=black,urlcolor=blue,filecolor=blue}

\begin{document}

\title{Polyakov loop and heavy quark entropy in strong magnetic fields from holographic black hole engineering}

\author{Renato Critelli}
\email{renato.critelli@usp.br}
\affiliation{Instituto de F\'{i}sica, Universidade de S\~{a}o Paulo, Rua do Mat\~{a}o, 1371, Butant\~{a}, CEP 05508-090, S\~{a}o Paulo, SP, Brazil}

\author{Romulo Rougemont}
\email{romulo@if.usp.br}
\affiliation{Instituto de F\'{i}sica, Universidade de S\~{a}o Paulo, Rua do Mat\~{a}o, 1371, Butant\~{a}, CEP 05508-090, S\~{a}o Paulo, SP, Brazil}

\author{Stefano I.~Finazzo}
\email{stefano@ift.unesp.br}
\affiliation{Instituto de F\'{i}sica Te\'orica, Universidade do Estado de S\~{a}o Paulo, Rua Dr. Bento T. Ferraz, 271, CEP 01140-070, S\~{a}o Paulo, SP, Brazil}

\author{Jorge Noronha}
\email{noronha@if.usp.br}
\affiliation{Instituto de F\'{i}sica, Universidade de S\~{a}o Paulo, Rua do Mat\~{a}o, 1371, Butant\~{a}, CEP 05508-090, S\~{a}o Paulo, SP, Brazil}

\begin{abstract}
We investigate the temperature and magnetic field dependence of the Polyakov loop and heavy quark entropy in a bottom-up Einstein-Maxwell-dilaton (EMD) holographic model for the strongly coupled quark-gluon plasma (QGP) that quantitatively matches lattice data for the $(2+1)$-flavor QCD equation of state at finite magnetic field and physical quark masses. We compare the holographic EMD model results for the Polyakov loop at zero and nonzero magnetic fields and the heavy quark entropy at vanishing magnetic field with the latest lattice data available for these observables and find good agreement for temperatures $T\gtrsim 150$ MeV and magnetic fields $eB\lesssim 1$ GeV$^2$. Predictions for the behavior of the heavy quark entropy at nonzero magnetic fields are made that could be readily tested on the lattice.
\end{abstract}



\maketitle


\section{Introduction}

The early stages of noncentral ultrarelativistic heavy ion collisions \cite{noncentralB1,noncentralB2,noncentralB3,noncentralB4,noncentralB5,noncentralB6} provide a way to produce extremely large magnetic fields of the order of $eB\sim 15m_\pi^2\sim 0.3$ GeV$^2$ at the top collision energies of the Large Hadron Collider. Such strong fields may have consequences for the transport and thermodynamic properties of the quark-gluon plasma (QGP) \cite{expQGP1,expQGP2,expQGP3,expQGP4,expQGP5,QGP,reviewQGP1,reviewQGP2} formed in later stages of heavy ion collisions and this possibility, in conjunction with the relevance of intense magnetic fields also in other environments such as the interior of magnetars \cite{magnetar} and the early Universe \cite{universe1,universe2}, have boosted the interest in the investigation of the properties of strongly interacting matter under the influence of strong magnetic fields \cite{reviewfiniteB1,reviewfiniteB2,reviewfiniteB3,reviewfiniteB4}. 

Moreover, it is well known by now that the QGP produced in heavy ion collisions behaves as an almost perfect, strongly coupled fluid close to the QCD crossover transition \cite{Aoki:2006we}, as evidenced, for instance, by the very small value of the shear viscosity over entropy density ratio $\eta/s\approx 0.095$ used in hydrodynamic simulations \cite{Ryu:2015vwa} that match experimental data. In fact, this small value is very close to the result $\eta/s=1/4\pi$ \cite{Policastro:2001yc,Buchel:2003tz,Kovtun:2004de} valid for a broad class of holographic gauge/gravity \cite{adscft1,adscft2,adscft3,adscft4} systems, and at least one order of magnitude below perturbative QCD estimates \cite{Arnold:2000dr,Arnold:2003zc}. This, along with the fact that first principles lattice QCD simulations suffer from technical difficulties to cope with nonequilibrium, real time observables \cite{Meyer:2011gj}, while holography can be straightforwardly employed to calculate retarded Green's functions in strongly coupled systems \cite{Son:2002sd,Herzog:2002pc,Gubser:2008sz,Skenderis:2008dg}, sparked the interest to use the gauge/gravity duality as a proxy to gain insights into transport properties of strongly correlated quantum fluids, such as the QGP and ultracold atomic systems \cite{adams,solana}.

While initially focused in some qualitative and seemingly universal features of strongly coupled fluids, recently \cite{Gubser:2008ny,Gubser:2008yx,DeWolfe:2010he,DeWolfe:2011ts,Gursoy:2007cb,Gursoy:2007er,Gursoy:2010fj,Finazzo:2014cna,Rougemont:2015wca,Rougemont:2015ona,Finazzo:2015xwa}, models defined within the holographic correspondence have also been applied in a more quantitative fashion to study some of the thermodynamic and transport properties of QCD-like plasmas. The reasoning behind such an approach may be dubbed as ``holographic black hole engineering'' in the sense that a nontrivial dilaton field $\phi$, which breaks conformal invariance in the infrared, is introduced in the gravity action with its bulk profile controlled by a potential $V(\phi)$ dynamically fixed in order for the holographic equation of state at zero magnetic field and chemical potentials match the corresponding lattice QCD data. In this way, the black hole solutions of the model are adequately engineered to emulate some equilibrium properties of the QGP without extra conserved charges or electromagnetic sources. 

One may extend such black hole geometries by further considering the addition of a Maxwell field $A_\mu$ to the Einstein-dilaton system, defining an Einstein-Maxwell-dilaton (EMD) holographic model where the coupling function between the Maxwell and dilaton fields, $f(\phi)$, may be dynamically fixed by matching some appropriate QCD susceptibility, again, at zero magnetic field and chemical potentials. Depending on the susceptibility used to fix the Maxwell-dilaton coupling, one seeds the bottom-up EMD setup with a minimum amount of phenomenological data required to holographically describe a QCD-like plasma at finite magnetic field and/or different chemical potentials. Then, different EMD settings may be used to make holographic predictions for  equilibrium and nonequilibrium properties of the strongly coupled QGP at finite magnetic field and/or chemical potentials.

Very recently, an anisotropic version of the EMD model at finite temperature and magnetic field (and zero chemical potentials) was proposed in Refs.\ \cite{Rougemont:2015oea,Finazzo:2016mhm}, where the holographic magnetic equation of state and the magnetic field dependence of the crossover temperature were found to be in quantitative agreement with lattice QCD data with $(2+1)$ flavors and physical quark masses \cite{Bali:2011qj,Bali:2014kia}. In Ref.\ \cite{Finazzo:2016mhm}, many transport coefficients associated with momentum diffusion were computed as functions of temperature and magnetic field. Indeed, the quantitative agreement found with the lattice equation of state at finite magnetic field and the fact that the holographic setting has nearly perfect fluidity naturally built in and allows for the calculation of real time retarded correlators  makes the magnetic EMD model \cite{Rougemont:2015oea,Finazzo:2016mhm} a very natural candidate to be employed to study properties of the strongly coupled QGP that are still beyond the reach of first principle lattice techniques. However, in order to further check the range of applicability of the holographic EMD model, it is important to consider other direct tests besides the equation of state at finite magnetic field, and, more importantly, to make further predictions for nontrivial observables accessible to lattice calculations.

In this work, we check this framework against another important test by comparing the holographic Polyakov loop \cite{Maldacena:1998im,Rey:1998ik,Rey:1998bq,Brandhuber:1998bs,Andreev:2009zk,Noronha:2009ud} calculated on top of the magnetic EMD black hole solutions with lattice QCD data for this thermodynamic observable at finite magnetic field \cite{Bruckmann:2013oba,Endrodi:2015oba}. We also compare our EMD model result for the heavy quark entropy at zero magnetic field with the latest lattice data available \cite{Bazavov:2016uvm} and make the first predictions in the literature for this observable at nonzero magnetic field.

We use in this work units where $\hbar = k_B = c = 1$ and a mostly plus metric signature.

\section{Einstein-Maxwell-dilaton holographic model}

The anisotropic EMD holographic model at finite temperature and magnetic field we use has been discussed in detail in Refs. \cite{Rougemont:2015oea,Finazzo:2016mhm}, including the numerics required to solve the set of coupled equations of motion for the EMD fields. In this section, we give a brief overview on how the model is built and how their results for the magnetic equation of state compare to the corresponding lattice QCD data from Ref.\ \cite{Bali:2014kia}. We refer the interested reader to consult Refs.\ \cite{Rougemont:2015oea,Finazzo:2016mhm} for the technical details.

The {\it bulk action} for the EMD model is given by
\begin{align}
S&=\frac{1}{16\pi G_5}\int d^5x\sqrt{-g}\left[R-\frac{(\partial_\mu\phi)^2}{2}-V(\phi) -\frac{f(\phi)F_{\mu\nu}^2}{4}\right],
\label{eq:EMDaction}
\end{align}
while the ansatz for EMD fields with a constant and uniform magnetic field $\vec{B}$ pointing in the $z$-direction is of the form,
\begin{align}
ds^2&=e^{2a(r)}\left[-h(r)dt^2+dz^2\right]+ e^{2c(r)}(dx^2+dy^2)+\frac{dr^2}{h(r)},\nonumber\\
\phi&=\phi(r), \quad F=dA=\frac{B}{\Lambda^2}\, dx\wedge dy,
\label{eq:EMDansatz}
\end{align}
where the boundary of the asymptotically AdS$_5$ space is at $r\to\infty$ and the black hole horizon is given by the largest root of $h(r_H)=0$. We set to unity the radius of the asymptotically AdS$_5$ space and $\Lambda$ is a scaling factor with dimension of mass used to express in units of MeV physical observables calculated on the gravity side of the holographic correspondence \cite{Rougemont:2015oea,Finazzo:2016mhm}. In order to numerically solve the equations of motion coming from Eqs. \eqref{eq:EMDaction} and \eqref{eq:EMDansatz}, one needs to rescale the bulk spacetime coordinates to specify numerical values for the radial coordinate $r$ and the functions $a(r)$, $c(r)$, and $h'(r)$ at the horizon; with this, one may Taylor-expand the background functions $a(r)$, $c(r)$, $h(r)$, and $\phi(r)$ around the horizon and fix the corresponding Taylor coefficients as on-shell functions of a pair of initial conditions, $(\phi_0,\mathcal{B})$, where $\phi_0$ is the value of the dilaton at the horizon and $\mathcal{B}$ is the value of the rescaled magnetic field in these {\it numerical spacetime coordinates} \cite{Rougemont:2015oea,Finazzo:2016mhm}. By choosing different values of the pair $(\phi_0,\mathcal{B})$ and numerically integrating the EMD equations of motion from the horizon up to the spacetime boundary, one obtains different black hole geometries, each one of them corresponding to some definite physical state in the quantum field gauge theory. On the other hand, the free parameters of the EMD model are dynamically fixed by solving the equations of motions with $\mathcal{B}=0$ (implying zero magnetic field, $B$) and requiring that the holographic equation of state and magnetic susceptibility at $B=0$ match the corresponding lattice QCD results with $(2+1)$ flavors and physical quark masses from Refs. \cite{Borsanyi:2013bia} and \cite{Bonati:2013vba}, respectively. This gives \cite{Finazzo:2016mhm},
\begin{align}
V(\phi)&=-12\cosh(0.63\phi)+0.65\phi^2-0.05\phi^4+0.003\phi^6,\nonumber\\
f(\phi)&=0.95\,\textrm{sech}(0.22\phi^2-0.15\phi-0.32),\nonumber\\
G_5&=0.46, \quad \Lambda=1058.83\,\textrm{MeV}.
\label{eq:EMDparameters}
\end{align}

\begin{figure}[htp!]
\center
\subfigure[]{\includegraphics[width=0.8\linewidth]{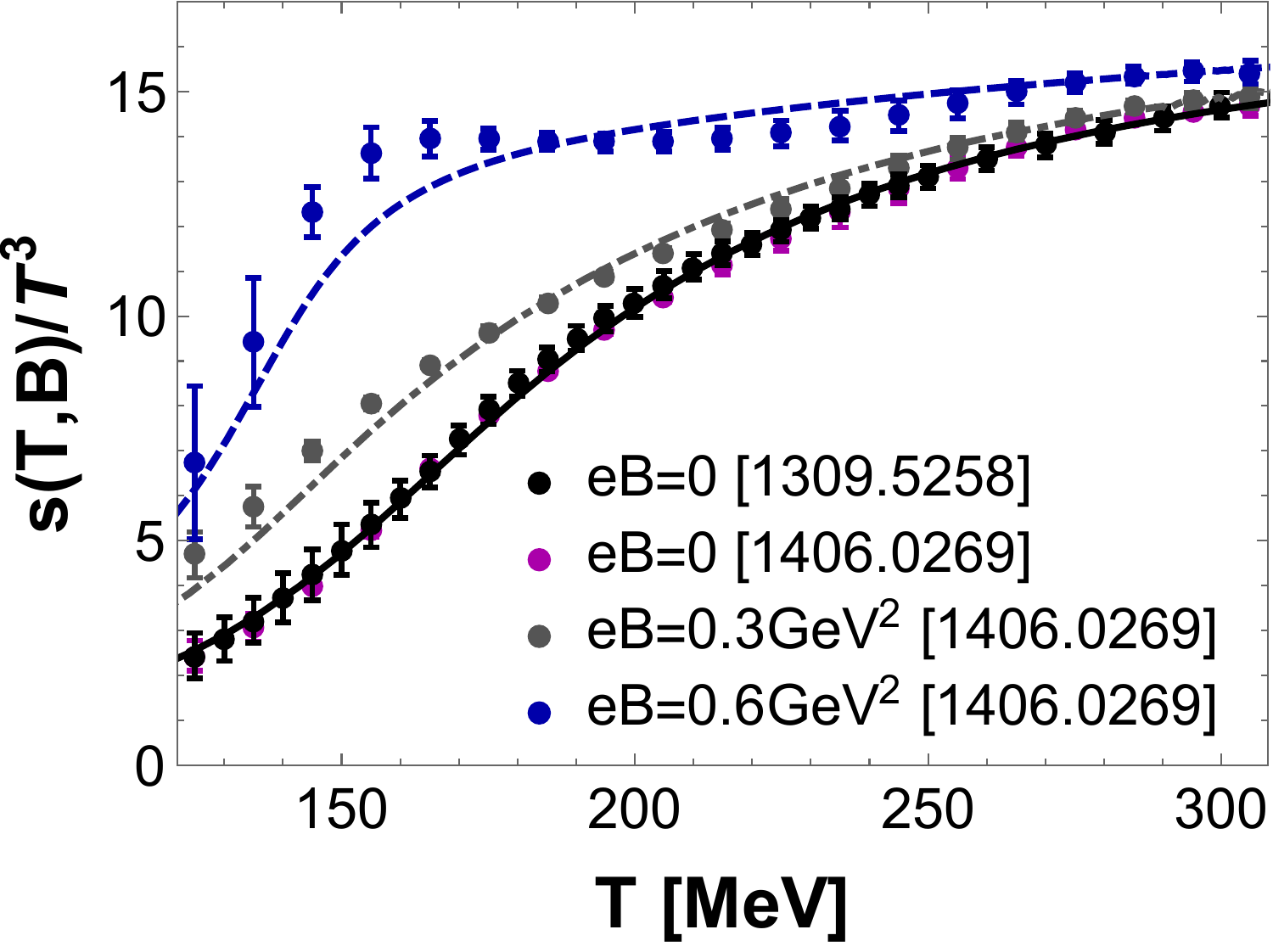}}
\qquad
\subfigure[]{\includegraphics[width=0.8\linewidth]{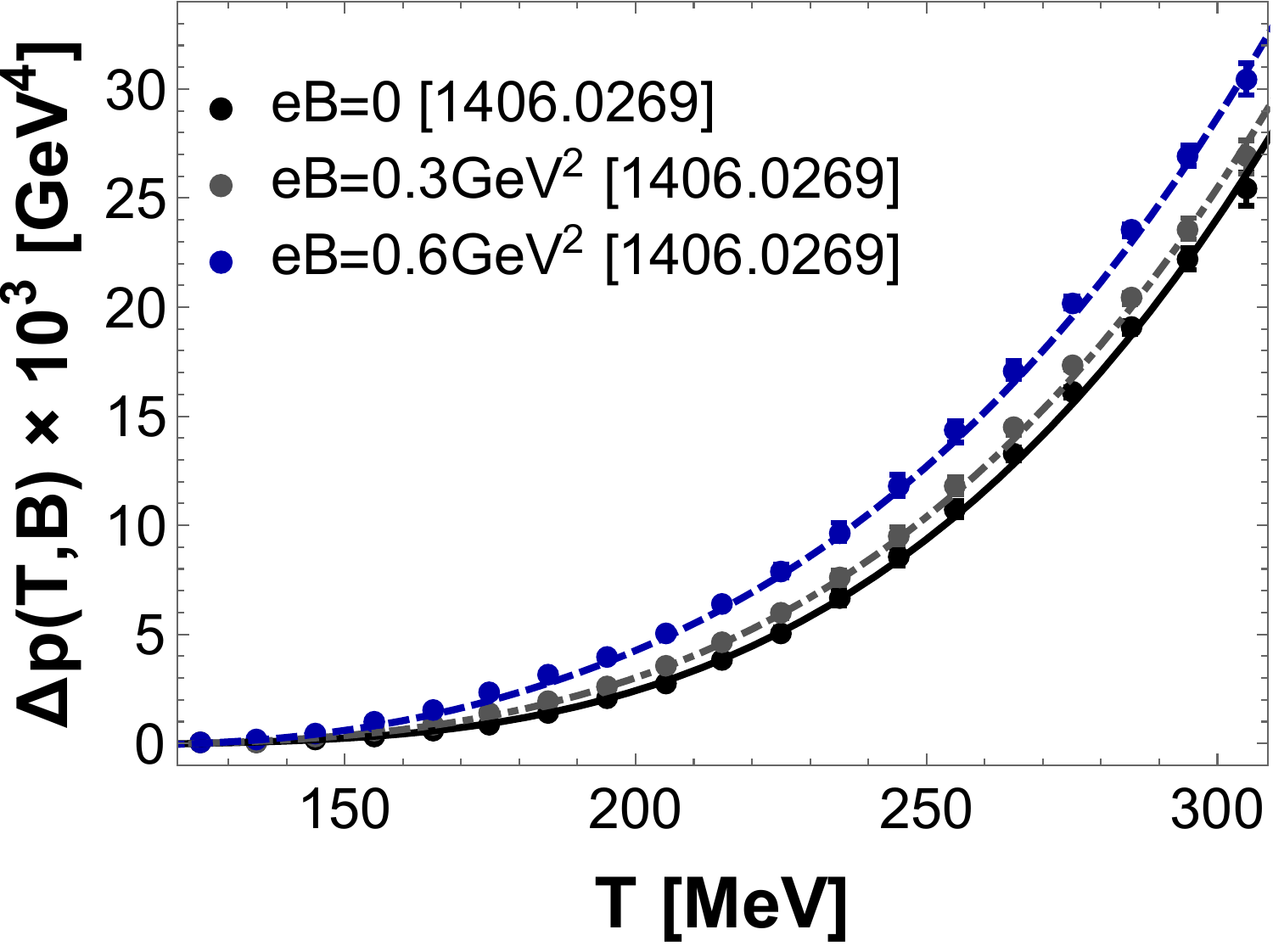}}
\qquad
\subfigure[]{\includegraphics[width=0.8\linewidth]{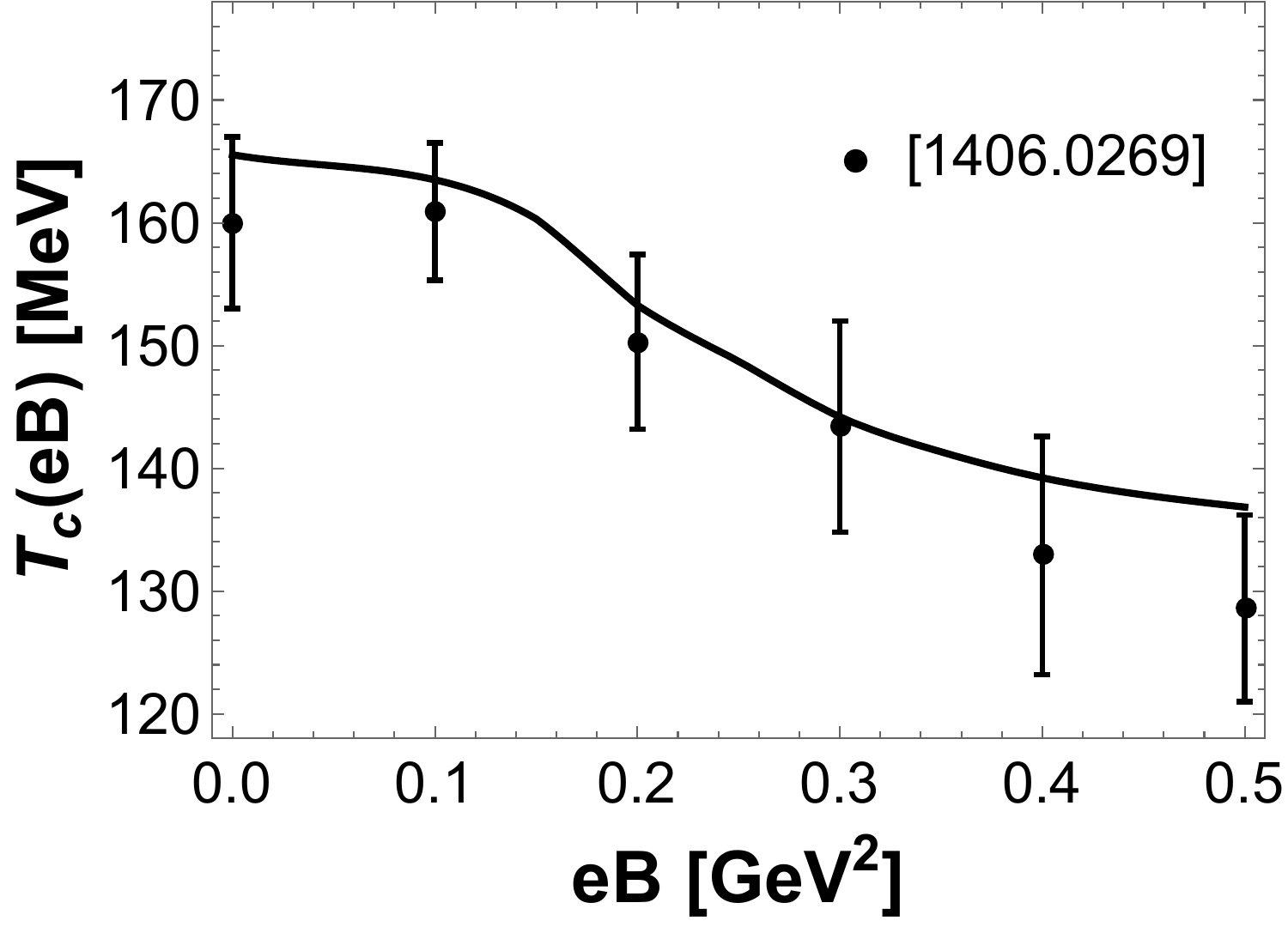}}
\caption{(Color online) Equation of state of the magnetic EMD model. (a) Normalized entropy density. (b) Pressure difference, $\Delta p(T,B)\equiv p(T,B)-p(T=125\textrm{MeV},B)$. (c) Crossover temperature extracted from the inflection point of the normalized entropy density.}
\label{fig:thermo}
\end{figure}

Since only the results for the equation of state and magnetic susceptibility at $B=0$ were used as phenomenological inputs to embed QCD information into the EMD holographic model, results for nonequilibrium transport coefficients at $B=0$ and $B\neq 0$ \cite{Finazzo:2016mhm} are legitimate holographic predictions of the present model, as well as the equation of state at nonzero magnetic field \cite{Finazzo:2016mhm}, which we show in Fig. \ref{fig:thermo} and compare to lattice data from Ref.\ \cite{Bali:2014kia}. One can see that the model provides a very good description of the lattice data.

\section{Polyakov loop and heavy quark entropy}

The holographic prescription for the calculation of the regularized free energy \cite{Bak:2007fk} of a single static heavy quark, $F_Q$, was studied in detail in Ref.\ \cite{Noronha:2009ud} and it involves the on-shell Nambu-Goto action for a straight string extending from the isolated heavy quark at the boundary up to the horizon in the interior of the bulk. The holographic formula for $F_Q$ in general backgrounds reads \cite{Noronha:2009ud,Finazzo:2013aoa,Finazzo:2014zga},
\begin{align}
F_Q&=\frac{\sqrt{\lambda_t}}{2\pi}\left[\int_{r_H}^{r_{\textrm{max}}}dr \left(\sqrt{-g^{(s)}_{tt}g^{(s)}_{rr}} -\sqrt{\textrm{Asy.}\left\{-g^{(s)}_{tt}g^{(s)}_{rr}\right\}}\right)\right.\nonumber\\
&\left.-\int_{\textrm{cte}}^{r_H}dr \sqrt{\textrm{Asy.}\left\{-g^{(s)}_{tt}g^{(s)}_{rr}\right\}}\right],
\label{eq:FQreg}
\end{align}
where $\sqrt{\lambda_t}=1/\alpha'=1/\ell_s^2$ is the effective t'Hooft coupling ($\ell_s$ is the fundamental string length), $g_{\mu\nu}^{(s)}=e^{\sqrt{2/3}\,\phi}g_{\mu\nu}$ is the string frame background metric \cite{Gursoy:2007cb,Gursoy:2007er,Gursoy:2010fj}, $\textrm{Asy.}\left\{-g^{(s)}_{tt}g^{(s)}_{rr}\right\}=e^{2r}$ is the asymptotic radial dependence of $-g^{(s)}_{tt}g^{(s)}_{rr}=e^{\sqrt{8/3}\,\phi(r)+2a(r)}$ in the ultraviolet limit $r=r_{\textrm{max}}\to\infty$, and $\textrm{cte}$ is an arbitrary constant corresponding to the choice of regularization scheme (note $\textrm{cte}\neq r_H$ in order for the regularization scheme employed in Eq.\ \eqref{eq:FQreg} be temperature independent - see \cite{Ewerz:2016zsx} for a recent discussion about the different regularization schemes). We must also remark that since our backgrounds support a nontrivial Maxwell field describing a constant magnetic field in the gauge theory, one could consider minimally coupling one string end point at a flavor brane near the boundary to the gauge field on top of it, as done, for instance, in Ref. \cite{Kiritsis:2011ha} in the case of finite mass quarks. However, in our calculations we consider infinitely heavy probe quarks, in which case the minimal coupling between the string and the gauge field is suppressed in the t'Hooft coupling relatively to the Nambu-Goto action (namely, the Nambu-Goto contribution is of order 1/2 in the t'Hooft coupling, while the minimal coupling term is of order 0). Consequently, for infinitely heavy probes, the minimal coupling between the strings and the gauge field does not contribute to the holographic calculation of the static heavy quark free energy in the classical gravity limit of the gauge/gravity duality, since the t'Hooft coupling is large in this limit. Moreover, the usual coupling between the dilaton field $\phi$ and the Ricci scalar $\mathcal{R}$ induced on the string worldsheet, which is of the form $\phi\mathcal{R}$, is also of order 0 in the t'Hooft coupling and, consequently, it is negligible in the aforementioned limit and may be ignored in the present calculations.

\begin{figure}[htp!]
\center
\subfigure[]{\includegraphics[width=1.0\linewidth]{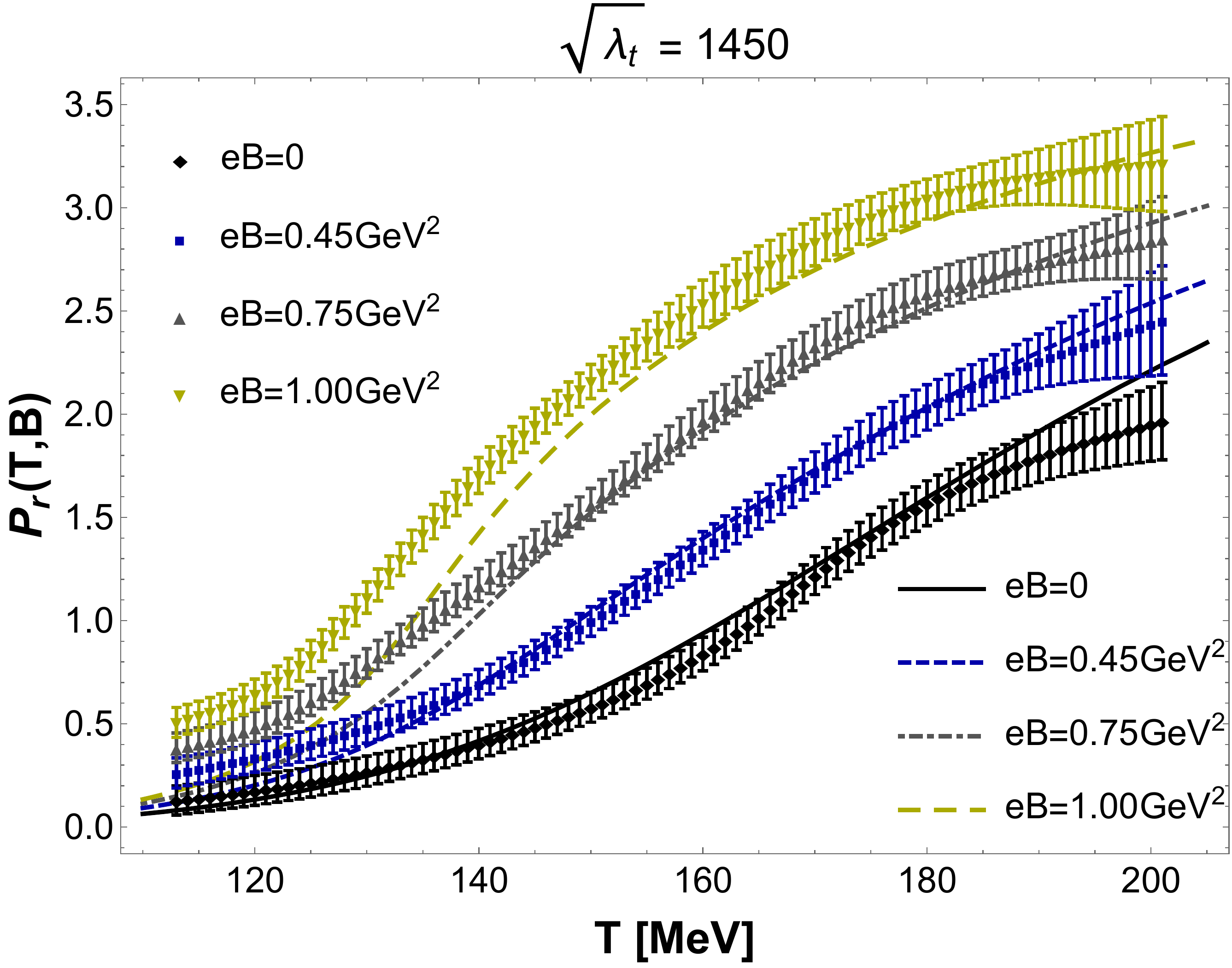}}
\qquad
\subfigure[]{\includegraphics[width=1.0\linewidth]{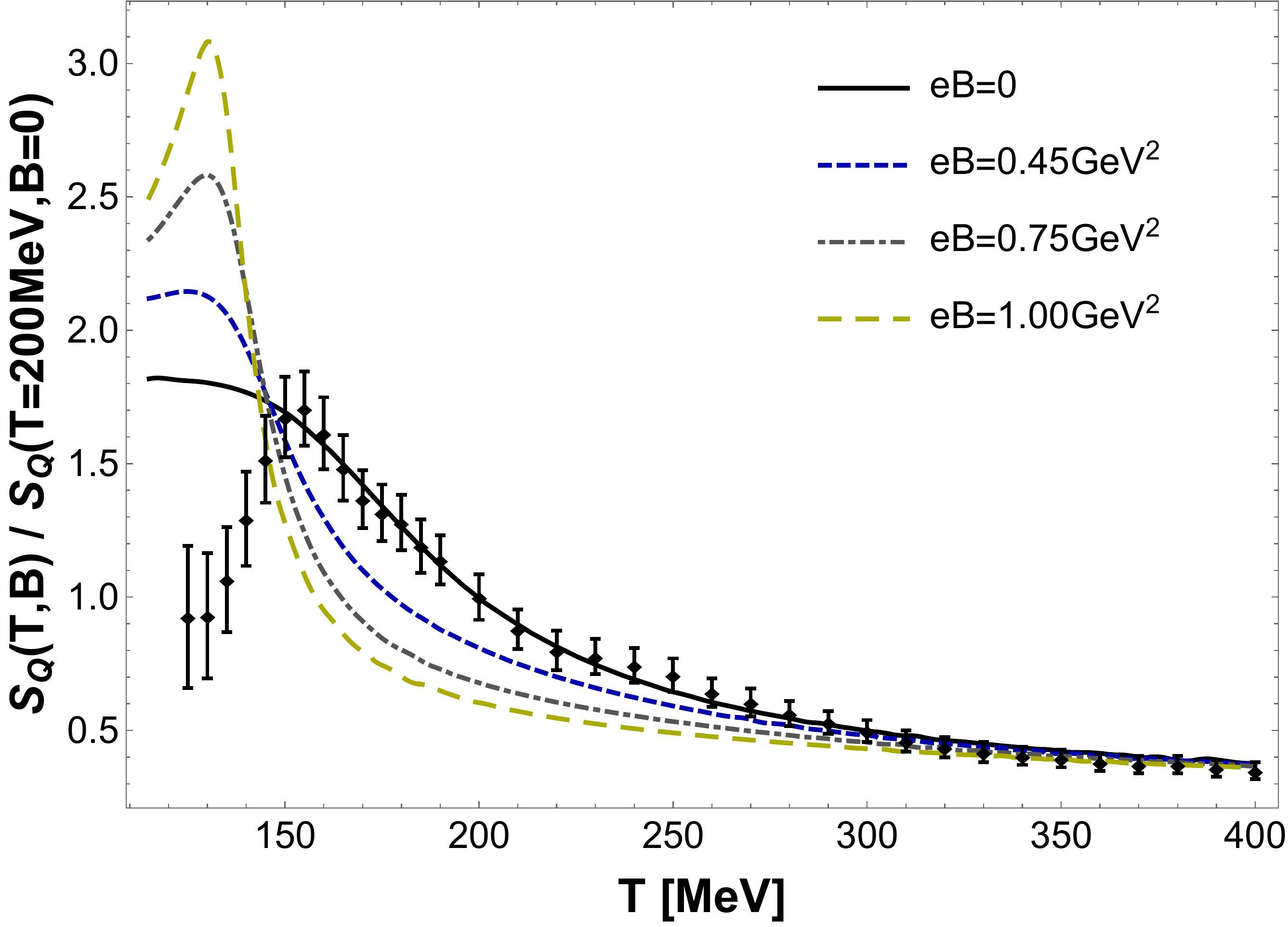}}
\caption{(Color online) (a) Renormalized Polyakov loop. Lattice data are taken from Refs.\ \cite{Bruckmann:2013oba,Endrodi:2015oba}. (b) Heavy quark entropy ratio. Lattice data with $B=0$ are taken from Ref.\ \cite{Bazavov:2016uvm}.}
\label{fig:poly}
\end{figure}

The absolute value of the expectation value of the Polyakov loop operator \cite{Polyakov:1978vu,'tHooft:1977hy,'tHooft:1979uj,Svetitsky:1982gs,McLerran:1980pk,McLerran:1981pb} is given by $P\equiv|\langle \hat{L}_P\rangle|=e^{-F_Q/T}$. We follow the same renormalization convention employed in Refs.\ \cite{Bruckmann:2013oba,Endrodi:2015oba} and define $F_Q^r(T,B)\equiv F_Q(T,B)-F_Q(T=162\textrm{MeV},B=0)$, such that $P_r(T,B)=e^{-F_Q^r(T,B)/T}$. The t'Hooft coupling remains as a free parameter coming from the Nambu-Goto action. By internal consistency of the holographic setting, it must be a large number, since it is inversely proportional to the square of $\ell_s$, which is taken to be very small compared to the radius of the asymptotically AdS$_5$ space in the classical gravity limit of the holographic correspondence. We show in Fig.\ \ref{fig:poly} (a) the holographic results for the renormalized Polyakov loop with $\sqrt{\lambda_t}=1450$ compared to lattice data at finite $B$ from Refs.\ \cite{Bruckmann:2013oba,Endrodi:2015oba}. One can see that the model is able to quantitatively describe the lattice results for the Polyakov loop at $B=0$ and that the agreement remains when the magnetic field is turned on up to $eB\lesssim 1$ GeV$^2$ for $T\gtrsim 150$ MeV.

The heavy quark entropy is given by $S_Q=-\partial F_Q/\partial T$. Previous holographic calculations of this quantity include the early work \cite{Andreev:2006nw} and Refs.\ \cite{Noronha:2010hb,Iatrakis:2015sua}. This quantity is particularly interesting in the holographic setting because the (zero temperature) regularization constant in $F_Q$ cancels after taking the derivative and the ratio of any two different values of $S_Q$ does not depend on the free parameter $\sqrt{\lambda_t}$ present in the holographic calculation of the Polyakov loop. Therefore, once the background is fixed, there are no free parameters to adjust in this calculation and the result is an unambiguous prediction of the holographic setup in equilibrium. In Fig.\ \ref{fig:poly} (b), we compare our holographic result for the heavy quark entropy ratio (with respect to the reference value at $S_Q(T=200\textrm{MeV},B=0)$) and the corresponding lattice data from Ref.\ \cite{Bazavov:2016uvm}, and also provide the first predictions for this ratio at nonzero $B$. One can see that the model gives a very good description of the $B=0$ data \cite{Bazavov:2016uvm} for $T\gtrsim150$ MeV. At high temperatures our results follow the approximate scaling $S_Q\sim 1/T^2$ suggested in \cite{Megias:2005ve}. At lower temperatures, the disagreement we find indicates that a different description involving hadronic states should be more appropriate, as discussed in \cite{Megias:2012kb,Bazavov:2013yv,Megias:2016bhk}.

Our model predicts that the heavy quark entropy decreases with increasing $B$ in a narrow region defined by $150 < T < 300 $ MeV, above which the curves computed with different values of $B$ approximately coalesce to a single curve at large temperatures. This feature can be readily checked on the lattice.


\section{Discussion}

The results in Figs. \ref{fig:thermo} and \ref{fig:poly} show that the magnetic EMD model is not only able to reproduce lattice results for the equation of state at nonzero values of the magnetic field, but also the results for the Polyakov loop above the pseudocritical crossover temperature, $T\gtrsim 150$ MeV, and magnetic fields $eB\lesssim 1$ GeV$^2$, with a reasonable choice for the t'Hooft coupling. As far as we know, this is the only model that is able to simultaneously match in a quantitative way lattice QCD results for both the equation of state and the Polyakov loop, at zero and nonzero magnetic fields. 

Furthermore, the holographic result for the heavy quark entropy ratio at vanishing magnetic field, which does not depend on the choice of any free parameter, agrees quantitatively with the latest lattice results available for $T\gtrsim 150$ MeV. These results constitute a highly nontrivial, empirical check of the holographic dictionary in a phenomenological bottom-up scenario engineered to describe the deconfined plasma phase of QCD. Taken together with the fact that the holographic setting has naturally built in the nearly perfect fluidity property of strongly correlated quantum fluids, which is a necessary condition for a {\it bona fide} description of the strongly coupled QGP, the results presented in this work give strong support for the use of the holographic EMD model in calculations for nonequilibrium, real time properties of the strongly coupled magnetized QGP, which are very difficult to investigate via first principle lattice techniques.

In the present work, we also made the first predictions for the behavior of the heavy quark entropy at nonzero magnetic fields, which could be tested against lattice simulations providing a further check of the phenomenological reliability of the holographic model described here.

We close this paper by making some considerations about the nature of the scalar field $\phi$ in the effective EMD action \eqref{eq:EMDaction}. We considered here that $\phi$ is the dilaton field, whereby the string and Einstein frame background metrics are different, with the former entering in the expression for the Nambu-Goto action used to calculate the holographic Polyakov loop and the heavy quark entropy, as discussed before. However, since our EMD model is a bottom-up construction, we do not know an explicit embedding of it into string theory and, consequently, we do not have a formal proof that $\phi$ is indeed the dilaton. In this regard, one could consider, alternatively to the viewpoint adopted here, that $\phi$ is some scalar field other than the dilaton. This a legitimate possibility as discussed, for instance, in Refs. \cite{Ewerz:2016zsx,DeWolfe:2009vs}. By following the reasoning discussed in these references, we also investigated what happens if we do not interpret the scalar field $\phi$ as the dilaton, in which case the metric in the string frame would be simply equal to the Einstein frame metric, i.e., $g_{\mu\nu}^{(s)}=g_{\mu\nu}$. In this case, we checked that the holographic results for both the Polyakov loop and the heavy quark entropy have nothing to do with the corresponding lattice QCD data, even at the qualitative level. Since the parameters \eqref{eq:EMDparameters} of the EMD action \eqref{eq:EMDaction} were dynamically fixed in Ref. \cite{Finazzo:2016mhm} by matching lattice QCD results for the equation of state and the magnetic susceptibility at $B = 0$, one could expect that this effective action should produce results compatible with QCD for some other physical observables, and what we concluded is that such expectation can only be fulfilled for the Polyakov loop and the heavy quark entropy if we interpret $\phi$ as the dilaton (or, at least, as a field which couples to the strings in the very same way as the dilaton does). Of course, this is not a theoretical proof that $\phi$ in our EMD action is the dilaton, but such interpretation seems to be phenomenologically useful for holographic QCD applications.\footnote{Such an interpretation may give rise to a further question regarding the fact that $\phi$ in the present EMD model is massive, while the dilaton field appearing in string theory is (usually) massless. However, there are processes such as supersymmetry breaking, which gives a mass to the dilaton field in string theory. In principle, another possibility could be that $\phi$ is a 5D massive Kaluza-Klein mode of the massless dilaton in 10D obtained after some specific 5D compactification. However, since our EMD model is a bottom-up construction, we cannot say at present which is the origin of the mass of the $\phi$ field in the effective action \eqref{eq:EMDaction}. This is a very interesting question, but also a very difficult one, which certainly goes beyond the scope of the present work.}\\


\begin{acknowledgments}
We thank G. Endrodi for making available to us lattice data for the Polyakov loop and H. Nastase for fruitful discussions. R.C. acknowledges financial support by Conselho Nacional de Desenvolvimento Cient\'{i}fico e Tecnol\'{o}gico (CNPq). R.R. acknowledges financial support by the S\~{a}o Paulo Research Foundation (FAPESP) under FAPESP grant number 2013/04036-0. S.I.F. was supported by FAPESP and Coordena\c{c}\~ao de Aperfei\c{c}oamento de Pessoal de N\'{i}vel Superior (CAPES) under FAPESP grant number 2015/00240-7. J.N. acknowledges financial support by FAPESP and CNPq.
\end{acknowledgments}


\end{document}